\documentclass[11pt,twoside]{article}
\usepackage{./asp2014}

\aspSuppressVolSlug
\resetcounters

\begin{document}

\title{Galactic Archaeology: 
Current Surveys and Instrumentation
}
\author{Rosemary F.~G.~Wyse
\affil{Johns Hopkins University, Baltimore, MD, USA; \email{wyse@jhu.edu}}}


\paperauthor{Rosemary F.~G.~Wyse}{wyse@jhu.edu}{}{Johns Hopkins University}{Department of Physics \& Astronomy}{Baltimore}{MD}{21218}{USA}

\begin{abstract}
I present an overview of the science goals and achievements of ongoing spectroscopic
surveys of individual stars in the nearby Universe.  I include a brief
discussion of the development of the field of Galactic Archaeology - using the fossil record
in old stars nearby to infer how our Galaxy evolved and place the Milky Way in cosmological context. 

\end{abstract}

\section{Introduction: The Fossil Record}

Galactic Archaeology is the study of the properties of old, low-mass
stars nearby, to unravel the evolution of the Milky Way, a typical
large disk galaxy. Low-mass stars (of mass similar to, or less than,
that of the Sun) have main-sequence lifetimes of at least
10$^{10}$\,yr. This is of order the age of the Universe, so that most
low-mass stars ever formed remain today. Indeed there are copious
numbers of stars nearby that have ages of order 10~Gyr; equating these
ages to lookback times (for the concordance cosmology) reveals that
these stars formed, and thus reflect conditions, at redshifts greater
than 2. These stars retain some memory of the initial early conditions
at which they form, and thus the early stages of galaxy evolution. The
chemical elemental abundances in the stellar photospheres of main
sequence stars are essentially conserved, {\it modulo} mass transfer in
close binary systems. Orbital angular momentum about the $z-$axis is an approximate
integral of the motion (exact in a time-independent axisymmetric
potential) and kinematic signatures can persist over many orbital
times. Studying low-mass stars nearby offers a complementary approach
to direct study of galaxies at high redshift: evolution of one galaxy
through time {\it versus} snapshots of different galaxies at different
times. Data from resolved stars allow the derivation of metallicity independent
of age and provide strong constraints on the stellar Initial Mass Function,
breaking degeneracies often encountered in the analysis of the integrated
light of galaxies. 

The information that can be extracted from the stellar fossil record
pertains to a range of physical processes, including 
\begin{itemize}
\item The merging
history of the galaxy, which is dependent on the nature of Dark Matter through the
power spectrum of primordial density fluctuations
\item The star-formation history - which may be quite different from the mass assembly history 
\item Chemical evolution, with associated constraints on stellar \lq feedback'
\item The relative importance of dissipative gas physics compared to dissipationless processes
\item Stellar Initial Mass Function at low and high redshift
\item Derivation of the density profile of the total mass from the stellar kinematics 
\item The evolution of dark matter substructure and baryonic substructure
\end{itemize}

\section{A Survey of Surveys}

There have been many stellar surveys which share the same `big picture' science
goals - to decipher the evolutionary history of the Milky Way - but which adopt a 
variety of approaches. They have targeted different components and/or
used different tracers such as clusters, main sequence turn-off stars,
red giant stars, blue horizontal stars and red clump stars.  The
surveys also differ in which phase-space coordinates can be studied:
2D or 3D spatial coordinates, 1D or 3D kinematics,  overall metallicity 
or detailed elemental abundances.  Ideally the survey would be capable
of analyzing the sample in as many properties as possible. The
quality and quantity of data also distinguish different surveys, and
experience has emphasized the need to understand, and minimize, both
systematic and random uncertainties. As capabilities have improved,
larger samples sizes have enabled analyses of parameters 
to go beyond simple means and dispersion to the full distribution functions - there is much physics in the detailed shapes, and in  \lq extreme objects'. 

I will give a brief survey of surveys, focusing on the disks of the Milky Way.

\subsection{Pioneering Precursors} 

At this conference we are discussing surveys of stars with sample
sizes in the millions. Early surveys - essentially those prior to the
advent of multi-object spectrographs in the mid 1980s - were limited
to samples of a few hundreds of stars, observed in the immediate solar
vicinity, or used globular clusters as more distant, intrinsically
luminous tracers (assuming their members were described by delta
functions in age and metallicity). The enduring lessons from these
surveys were not always the immediate ones. The pioneering study by
\citet{Eggen62}  (hereafter ELS) demonstrated how analysis of
kinematics and chemical abundances could provide insight into how the
Milky Way Galaxy formed. That study analyzed data for a sample of 221
nearby main-sequence stars, derived photometric metallicity estimates
from UV-excess and used that as a proxy for age. They noted a strong
anti-correlation between each of orbital angular momentum and vertical
velocity and metallicity, which they famously interpreted as evidence
for rapid collapse of a monolithic star-forming gas cloud from a
short-lived halo phase to a centrifugally supported disk. This
conclusion was challenged by \citet{SZ78} (hereafter SZ) based on inferences
from a sample of $\sim 50$ Galactic globular clusters, with
metallicity indicators (mostly taken from the literature) and
photometry below the horizontal branch. These data showed no evidence
for a metallicity gradient in the outer halo (defined by Galactocentric distances greater than 12~kpc), not obviously consistent with the strong anti-correlation between kinematics and
metallicity found in the field stars. Further, SZ  found
that the morphology of the horizontal branch, as measured by the
relative numbers of stars redward and blueward of the RR~Lyrae gap,
varied within the population of outer-halo globulars, even at fixed metallicity. The inner-halo clusters did not show such an effect. The \lq second parameter' (metallicity being the first) affecting the colors of horizontal branch stars was taken to be age, with the observed range implying an age spread of several Gyr for the outer halo. This is at odds with the rapid collapse of ELS. Searle \& Zinn proposed their data implied that the globular clusters of the outer halo formed in \lq transient protogalactic fragments' that were accreted \lq some time after collapse of its central regions'. This of course is reminiscent of the later infall of satellite galaxies predicted in $\Lambda$CDM models.   

It was clear from their apparently contradictory conclusions that it
was not possible that both ELS and SZ could be correct.  We now have a
better understanding of the biases that can arise in proper-motion
selected samples such as that used by ELS \citep[see, for example,][]{Bond70,CLL89}. We also now have ages of globular clusters that
are derived from fitting of the main-sequence turnoff and
spectroscopic metallicities; these have revealed that while age indeed
is the most plausible `second parameter', all metal-poor globular
clusters are old, while the metal-rich ones have a range of ages and
have more disk-like kinematics \citep[e.g.][]{Dotter11,Leaman13}. Several of each of the younger and
older clusters are associated with the Sagittarius dwarf \citep{Ibata97}, 
validating that some part of the outer halo cluster
system is accreted.  The bulk of the stars in the halo are indeed old,
although that does not mean that they formed {\it in the Galaxy\/} a
long time ago.  The moral of the story is that good understanding and
control of selection biases are critical, as are large enough samples
with accurate and precise enough data.

The limitations of small sample sizes and large uncertainties in
metallicity estimates are manifest in the ease with which \citet{Tinsley75} could fit
the cumulative metallicity distribution of nearby stars (which shows the famous \lq G-dwarf problem' - the lack of metal-poor long-lived stars compared to predictions of the Simple Model) with very different
models.  Cumulative distributions, appropriate for such limited data,
do not provide sufficient discrimination among models.

\subsection{Moving Beyond the Solar Neighborhood}

The advent of scanning machines for automated measurement of positions
and photometry from photographic plates facilitated the analysis of
wide-area star counts to faint magnitudes. The density law at the
South Galactic Pole, based on distances from photometric parallax,
including models of the vertical metallicity gradients, was shown by
\citet{GR83} to require an additional component to the standard thin
disk plus spheroid of existing models \citep[e.g.][]{BS80}. This \lq
thick disk' identified by Gilmore \& Reid was proposed by them to be
part of the stellar halo, albeit \lq Intermediate Population
II'. The multivariate dependence of the stellar luminosity function led to inevitable arguments about the robustness of this star-count analysis. Subsequent studies characterized the thick disk as distinct in
kinematics and metallicity from both the stellar halo and (old) thin disk \citep[e.g.][]{WG86,RF89,CLL89}. Not only the field stars were better described by models including a distinct third component, improved data for the globular cluster system of the Milky Way - long taken as tracing the stellar halo, as noted above - showed that they too showed a distinct \lq disk' component \citep{Zinn85}. 
The stars of the thick disk were found to be predominantly old and plausibly trace the earliest phase of disk formation \citep{JW83}, perhaps even providing a self-consistent solution to the \lq G-dwarf problem' \citep{GW86}.

\subsubsection{Near and Far}

The advent of multi-object fiber spectrographs, using first manual
plug-plate technology then automated positioners, opened the way for
spectroscopic metallicities and line-of-sight velocities for large
samples of stars, with the early generations of MOS in the 1980s
providing simultaneous spectra for around one hundred stars within a
field-of-view of a degree or so.  A joint analysis of the kinematics
and metallicity distributions of a sample of F/G main sequence stars
in the local solar neighborhood with a corresponding sample of stars
several kiloparsec away in the thin disk/thick disk interface, allows
decomposition into thick and thin disks where the distributions
overlap. Such an analysis, with sample sizes still only in the
hundreds in each line-of-sight \citep{WG95}, allowed the identification of
metal-poor thin disk stars and metal-rich thick disk stars, with the
assignment of a star to a given disk component being based on
kinematics and location within the Galaxy. The extents of the
metallicity distributions in these regimes provide important 
constraints on models of disk evolution
\citep[e.g.][]{WG95,Snaith15}. The joint distribution of colors and
spectroscopic metallicities for statistically significant samples of F/G main
sequence thick-disk stars allows an estimate of the \lq turnoff' age to
be made. Such an analysis consistently shows the thick disk to be 
predominantly old, with typical ages in the range of 10-12\,Gyr - the same as 47 Tuc, the prototypical \lq thick disk' globular cluster.

\subsection{Back Nearby, Larger Samples}

Wide-field imaging in narrow- and intermediate-band filters, such as
the Str\"omgren system, has the potential to provide very large samples
of stars with photometric metallicity and age estimates
\citep[][]{Strom87}. The Geneva-Copenhagen Survey
\citep[][]{Nord04,Cas11} of local thin disk stars ($\sim 14,000$ F/G
stars within $\sim 40$~pc) with Str\"omgren photometry, parallaxes
from Hipparcos, proper motions and line-of-sight velocities (hence
full 3D space motions) have revealed the wealth of information in
joint age-kinematics-metallicity analyses. The large scatter in the
age-metallicity relationship hinted at in earlier samples \citep[e.g.][discussed below]{Wall62} was
definitively confirmed, as was the lack of change with time in the mean
metallicity of the local disk. These data demonstrated the existence of stars with disk-like kinematics over the range of lookback times reaching almost to the cosmic \lq dark ages', with many old stars nearby. The large sample size also revealed details of substructures in the disk that are defined by kinematics, moving coherently, but are unlike \lq moving groups' in that they have a large internal range of age and metallicity. These signatures are best explained by internal resonant interactions between field disk stars and spiral arms and/or the bar. 

\subsection{Substructure}

Larger samples, with well-understood errors, allow the identification
and characterization of substructure in multi-dimensional kinematic-chemical-age phase space.
Imaging surveys play a crucial role, in both defining the morphology
in coordinate space and supplying targets for follow-up
spectroscopy. The \lq Field of Streams' in the distribution of faint
turnoff stars isolated by \citet{Bel06} in the Sloan Digital Sky
Survey (SDSS) imaging data was critical in determining the extent of
larger-scale substructure in the field stars of the Galaxy \citep[see also][]{New02} and to the
discovery of \lq ultra-faint' dwarf galaxies (although the numbers of newly discovered faint satellite galaxies still are
not sufficient for easy compatibility with the predictions of
$\Lambda$CDM models). The streams, tracing substructure that has yet
to mix dynamically, are dominated by tidal debris from the Sagittarius
dwarf galaxy, which was originally discovered as a moving group
\citep{IGI95} in a MOS survey of the Galactic bulge.  The \lq stream' structure closer to the plane of the
Galactic disk, in the Monoceros line-of-sight, is most likely composed of stars that belong to the Galactic disk \citep[e.g.][]{Xu15} rather than to a disrupted galaxy. 

Structure in kinematic phase space survives longer than does structure
in coordinate space, with structure in chemical space the most
persistent. Multi-object spectroscopy is required to characterize
substructure -- and indeed to determine the underlying
structure. Surveys should be designed to sample all expected \lq
interesting' scales, while still being able to make serendipitous
discoveries. This requirement leads to sample sizes of at least ten thousand
\citep[e.g.][]{apoorva13}. The dedicated survey mode, pioneered by
the Sloan Digital Sky Survey, has allowed total sample sizes in the hundreds
of thousand. 

\subsection{Dedicated Survey Mode} 

\begin{figure}[t]
 \centering
\includegraphics[width=12cm]{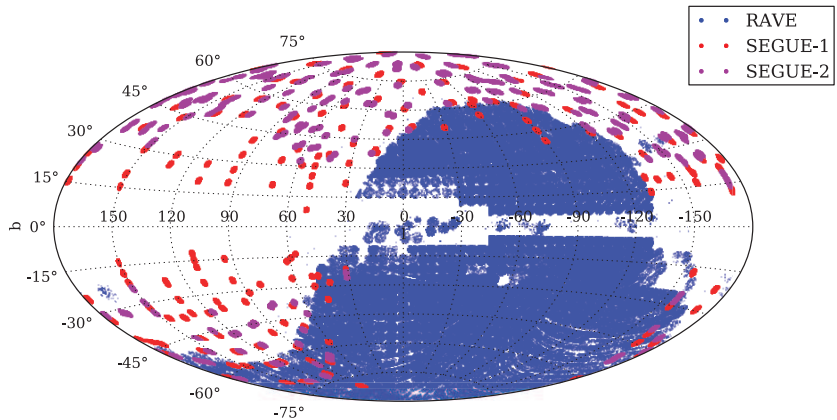}%
  \caption{ Footprints on the sky, in Galactic coordinates, of the SDSS SEGUE-1 (sparse-sampled, orange fields) and SEGUE-2 (sparse-sampled, high-latitude purple fields) surveys, together with that of the RAVE survey (densely sampled, blue fields).  Figure courtesy of Doroth\'ee Brauer.}
  \label{fig:RAVE_SEGUE}
\end{figure}

The Sloan Digital Sky Survey transformed the sociology of much of
astrophysics away from PI-led observing programs to survey
science. The initial SDSS spectroscopic survey observed stars only as
calibrators for the galaxy redshift survey.  The first SDSS dedicated
stellar spectroscopic survey, SEGUE \citep[][red fields in
Fig.~\ref{fig:RAVE_SEGUE}]{Yan09}, utilized these low-resolution
spectrographs ($ R \sim 1800$) with $\sim 600$ multiplexing to obtain
wide wavelength coverage spectra of $\sim 240,000$ stars in several
target spectral types and evolutionary phases (selected from the SDSS imaging data), including G
dwarfs, Blue Horizontal Branch stars and K Giants. The fields were
chosen to sparse-sample the sky, probing the large-scale (tens of kpc)
structure of the main stellar components of the Galaxy. The
pencil-beams of SEGUE-2 \citep[][purple fields in
Fig.~\ref{fig:RAVE_SEGUE}]{Eis11} were selected to probe high-latitude
substructure.

The spectra from the SDSS surveys provide line-of-sight velocities to
around $~10$~km/s. The values of the stellar atmospheric parameters are
obtained through a dedicated pipeline \citep{SSP08}, developed for the broad range of targets and giving
metallicity estimates to $\sim 0.2$~dex, and for high signal-to-noise spectra, an estimate of the
relative abundance of the \lq alpha-elements' (created by successive
addition of helium nuclei - alpha particles - in the interiors of
massive stars) to iron to similar accuracy and precision
\citep{SSP11}. The SEGUE data for $\sim 7,000$ main-sequence turnoff
stars at distances from the Sun of several kiloparsec were used to characterize further
the thick and thin disks, determining the large-scale radial metallicity gradient
as a function of vertical distance from the mid-plane
\citep{cheng12a}. Those authors found that radial gradient to flatten
with height, such that the thick disk has approximately constant mean
metallicity, as a function of radius.

The RAVE survey of $\sim 500,000$ bright stars is described in more
detail in this volume by Georges Kordopatis \citep[see
also][]{RAVE06,RAVE13}. On a philosophical/sociological note, the RAVE survey was supported by the
institutions and personal research grants of the participants and took
advantage of the opportunity offered by the withdrawal of government
support for the UK Schmidt, part of the worldwide transferral of
resources to larger (and larger) facilities. It is clear that 4\,m class
telescopes - and smaller - still have a critical role to play, particularly  in 
spectroscopic surveys.  The 6~degree field-of-view of the UK Schmidt
telescope and the $\sim 100$ multiplexing capability of the 6dF
spectrograph match well with the surface density of bright ($I < 12$)
stars across much of the sky. Further, moderate resolution (R $\sim
7000$), reasonable signal-to-noise spectra of such stars can be
obtained in around one hour, allowing a large sample to be studied in
a reasonable time (several years). The almost contiguous sampling on
the sky contrasts with the approach of the SEGUE surveys (as shown in
Fig.~\ref{fig:RAVE_SEGUE}) and allows both small-scale and large-scale
gradients/substructure to be studied. A selection of the major
scientific results is given in Georges Kordopatis' contribution,
including the complexity in the velocity distribution function of the
thin disk. The compression/rarefaction pattern of the vertical
velocities has recently been shown to be reflected in the stellar
density distribution, manifest in the SDSS imaging data. The star
counts of main-sequence stars show asymmetries above and below the
nominal Galactic Plane ($b = 0^{\rm{o}}$), consistent with a toy model in
which the plane of the disk is actually offset, alternating up and down at different radial ranges, by around 100\,pc, i.e.~roughly a third of the vertical scale-height of the thin disk \citep{Xu15}. 
This lends support to the interpretation  of low-latitude features such as the Monoceros Stream/Ring as being simply structures within the disk \citep{Xu15}, rather than being tidal debris from accreted satellite systems.  

I will return briefly to the SDSS-III APOGEE survey \citep{Eis11} below; this is described in more detail in Carlos Allende Prieto's contribution to this conference. The LAMOST survey was presented in the conference by Yongheng Zhao. 

\subsection{Elemental Abundances: Beyond Metallicity}

Different chemical elements are produced through the evolution of
stars of different (main sequence) masses and hence ejected into the
interstellar medium on different timescales. The elemental abundance
distributions therefore contain much more information about the
star-formation history and the stellar Initial Mass Function than does
the \lq metallicity' distribution \citep[e.g.][]{Tin79,GW98,Matt01}. In
particular, the alpha-elements 
are produced by short-lived massive stars that end
their lives in core-collapse supernovae, with the ejected mass of such
elements being a function of the mass of the progenitor
star. Core-collapse supernovae also eject a fixed, relatively small
mass of iron and the mass-averaged yields over the range of progenitor
stellar masses (above $\sim 8$~M$_\odot$) enrich the interstellar
medium with \lq alpha-enhanced' (i.e.~the ratio of alpha-elements to iron is greater than than in the Sun) gas, [$\alpha$/Fe] $\sim +0.3$ for
typical yields and typical massive-star IMF slope.  Stars that form during the
early stages of a star-formation event, when the enrichment is
dominated by core-collapse (Type II) supernovae, will therefore form from gas which is alpha-enhanced, and hence will have
photospheric abundances that are alpha-enhanced.

\begin{figure}
 \centering
\includegraphics[width=8cm,angle=270]{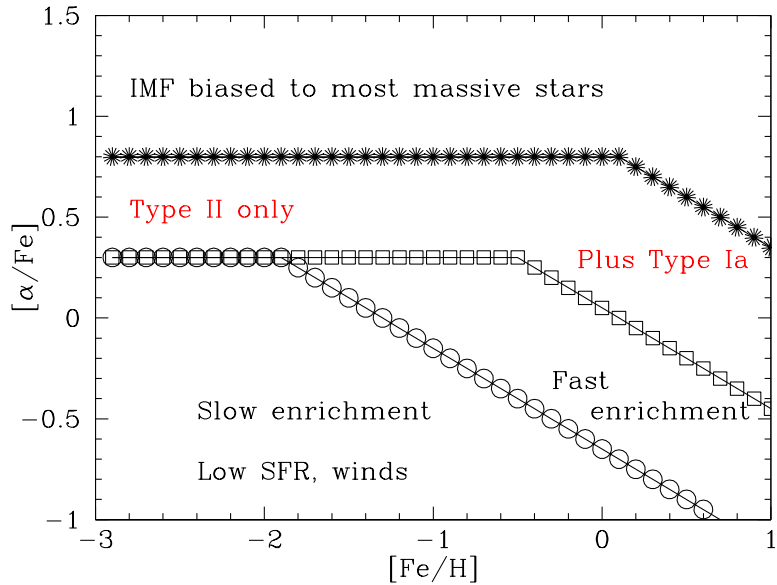}   
  \caption{Schematic of the expected pattern of elemental abundances for a self-enriching, well-mixed star-forming system, indicating the turndown from the \lq Type II plateau' at the time (measured from the onset of star formation) at which Type Ia supernovae have contributed significant iron to the interstellar medium. The iron abundance corresponding to this time depends on the past star-formation rate. The value of the \lq Type II Plateau' depends on the massive-star IMF due to the dependence of the yields on progenitor mass. The observed lack of scatter in real data implies little room for IMF variations (plus good mixing).}
  \label{fig:alpha}
\end{figure}

Type Ia supernovae, resulting from accretion onto a white
dwarf from a companion (either a non-degenerate binary companion or a
second white dwarf) that increases its mass to above the Chandrasekhar
limit, are delayed with respect to the core-collapse supernovae and
continue to occur on longer timescales (model-dependent, but
continuing many Gyr after birth of the progenitors). These also
produce a fixed mass of iron per event, but the explosive
nucleosynthesis of these events leads to an ejected mass in iron that is approximately ten times that
ejected by a core-collapse supernova, and little ejected mass in alpha-elements. Stars that form later in a
star-formation event, after the interstellar medium has been enriched
by iron-rich material from Type Ia supernovae, will have photospheric
abundances that are less alpha-enhanced than those formed earlier
(assuming a self-enriching system with no flows). The star-formation
history is thus encoded in the pattern of elemental abundances.
Elemental abundances are more straightforward to derive than are
estimates of stellar ages, and thus normally the iron abundance is
used as a proxy for age - albeit that there is known to be significant
scatter, of uncertain origin, in the local age-iron abundance 
relationship \citep{Nord04,Cas11}. 

A schematic plot of the predicted pattern of elemental abundances is
shown in Fig.~\ref{fig:alpha} \citep[based on an earlier figure in][]{WG93}. Core-collapse supernovae all explode within a short time after the onset of star-formation ($\sim 40$~Myr) and with good mixing in the interstellar medium the next generation of stars to form will be enriched with a (massive-star) IMF-average yield of metals. An invariant IMF will produce a \lq  plateau' in [$\alpha$/Fe] at
lowest values of [Fe/H] with the value of the plateau reflecting the distribution of stellar mass for
the progenitors of core-collapse supernovae, being higher for a
massive-star IMF that favors the most massive stars \citep[see,
e.g.][for a discussion of how one may exploit this dependence, plus the very low  amplitude of observed scatter, to
constrain IMF variations]{WG92}. The downturn at higher values of
[Fe/H], due to the addition of iron (but low ejected mass in alpha-elements) from Type Ia supernovae, occurs at
a fixed time but at an iron abundance that depends on the past
star-formation and enrichment rate.

\subsubsection{The Solar Neighborhood} 

\begin{figure}
 \centering 
\includegraphics[width=11cm]{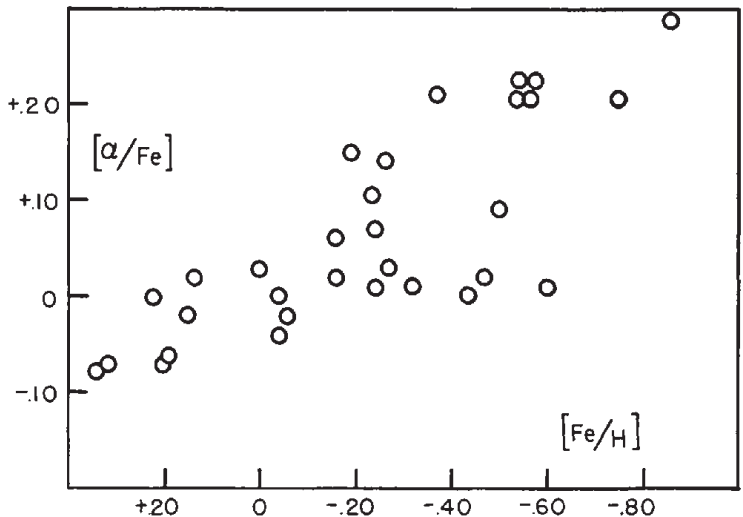}%
  \caption{Elemental abundances for a sample of around 30 field G stars, from the pioneering study of \citet{Wall62} (\copyright\  AAS; reproduced with permission). Note the reversal of the $x-$axis compared to modern convention.}
  \label{fig:wall}
\end{figure}

\citet{Wall62} carried out a pioneering survey of nearby (bright)
field G stars, obtaining elemental abundances from a curve-of-growth
analysis. As may be seen from Fig.~\ref{fig:wall}, he indeed found a 
population of more metal-poor stars that have enhanced
alpha-to-iron ratios, relative to the solar value ([$\alpha$/Fe]$ > 0$ is \lq enhanced'),
consistent with forming during the (short-duration) epoch when the
enrichment was dominated by core-collapse (Type~II) supernovae. The 
stars in this sample are bright enough to have trigonometric parallaxes, and
comparison with isochrones showed that there exist metal-rich old
stars, a result which is now  well-established.  The distribution in [$\alpha$/Fe] at
low values of iron abundance, showing two sequences, is also now a
robust result, as we return to below. In another precursor to modern
results, Wallerstein investigated the kinematics of the stars and
found that [$\alpha$/Fe] was a better predictor of orbital angular
momentum than was [Fe/H]: the \lq high-alpha' metal-poor stars are on
more eccentric orbits than are the \lq low-alpha' metal-poor stars.

Indeed several recent surveys of local stars with elemental abundances
derived from high-resolution spectra
\citep[e.g.][]{Fuhr11,Adi13,Bensby14} have confirmed these early results, finding a clear separation
into two sequences in the [$\alpha$/Fe], [Fe/H] plane, particularly at iron
abundances below the solar value. Again, the two sequences show different kinematics and age distributions and are usually associated with the thick disk (\lq high-alpha' sequence)
and the thin disk (\lq low-alpha' sequence). 

\subsubsection{Non-local samples}

The separation in the elemental abundance plane was however challenged by 
\citet{Bovy12a, Bovy12b} 
who analyzed
the distributions of elemental abundances derived from the SDSS SEGUE
data (low-resolution spectra), probing distances of several kiloparsec
from the Sun, and proposed that there was instead a continuum of \lq
mono-abundance populations' \citep[see also][]{JEN87}. This was difficult to reconcile
with the results from the higher resolution, higher signal-to-noise
data for the local stars.  There was an obvious need for a sample of
more distant stars with more precise elemental abundances, derived from
high-resolution, high signal-to-noise spectra, to match the local samples. 

\begin{figure}[t]
 \centering 
\includegraphics[width=13cm]{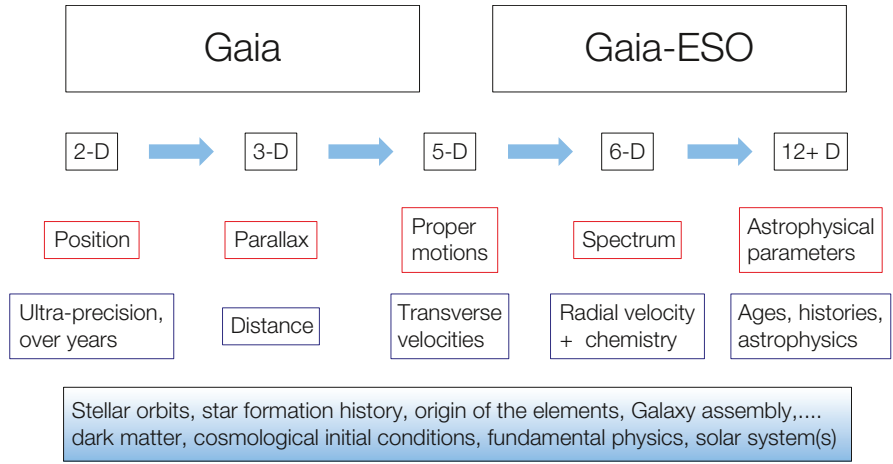}%
  \caption{Illustration of the gain in physical probes with the combination of astrometry from the Gaia satellite and wide-area spectroscopic surveys, in this case the Gaia-ESO survey \citep{GES12}.} 
  \label{fig:Gaia}
\end{figure}

Further, the selection and subsequent successful launch of the Gaia
astrometric satellite strengthened the scientific case for a
spectroscopic survey for (the fainter) stars for which Gaia would
provide astrometric (and photometric) data only. The complementarity of  ground-based spectroscopic surveys and space-based astrometric surveys is illustrated in Fig.~\ref{fig:Gaia}, taken from \citet{GES12}. This led to the
Gaia-ESO survey \citep{GES12}, allocated 300 nights on the VLT (over 5
years, starting January 2012) to obtain high-resolution ($R > 16,000$)
spectra of an intended sample of $10^5$ faint field stars (typically 
$r \sim 18$) in all major components of the Milky Way, plus a sample of
brighter stars in (open) star clusters. The field star targets are mostly F/G dwarfs and K giants, selected from VISTA imaging. Most of the spectra are obtained using FLAMES/GIRAFFE, with a multiplexing of 110 and field-of-view of 25~arcmin.

\begin{figure}[t]
 \centering 
\includegraphics[width=10cm]{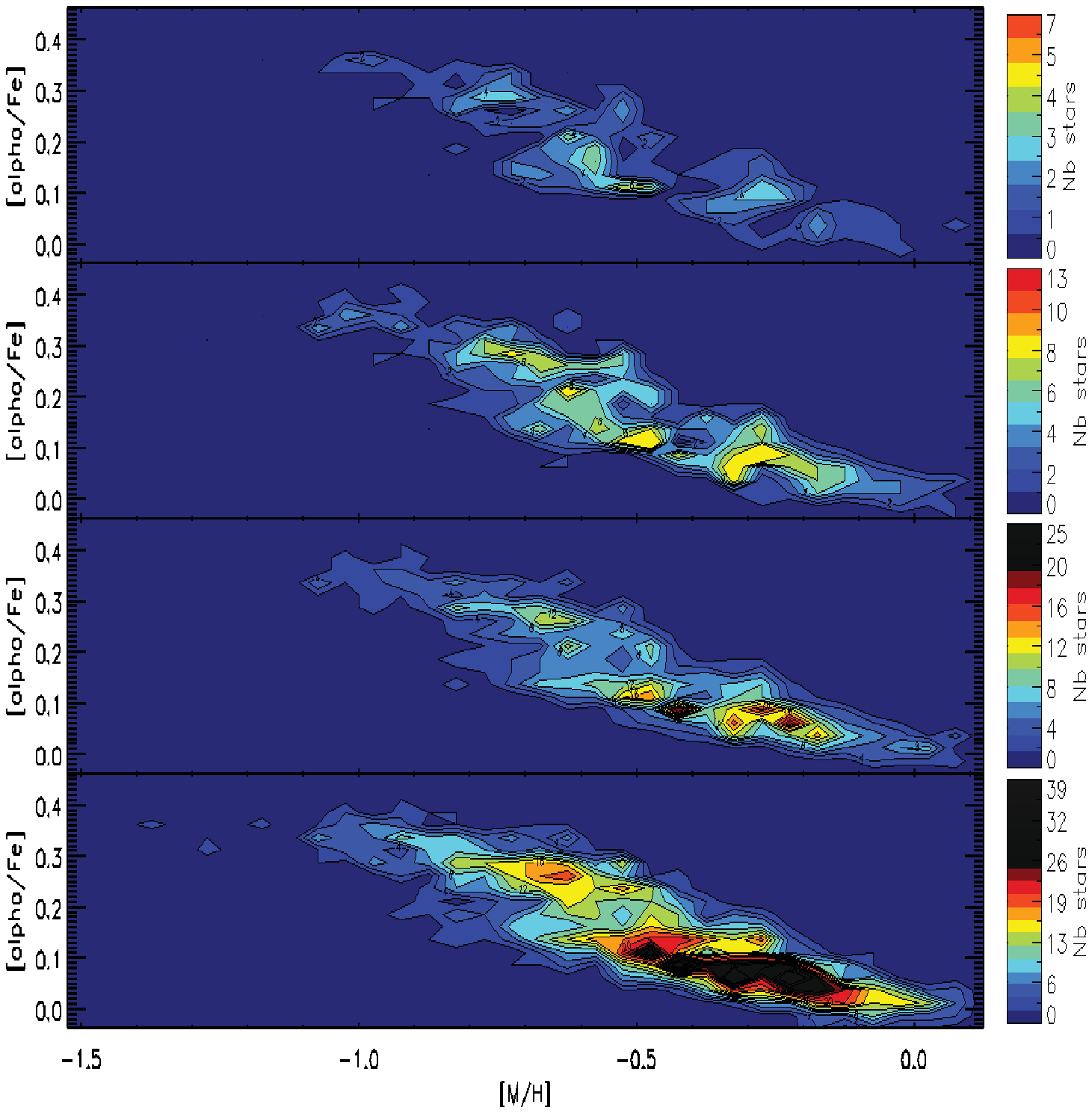}%
  \caption{Elemental abundances from the first internal data release
  of the Gaia-ESO Survey \citep{GES14}. Errors increase top to
  bottom. The top panel shows the distribution for the $\sim 200$
  stars with errors on [$\alpha$/Fe] below 0.03~dex and errors on
  metallicity ([M/H]) below 0.07~dex. The sample size increases as the
  error cut is relaxed - the third panel shows $\sim 1000$ stars with maximum errors in [$\alpha$/Fe] and in [M/H] of 0.05 dex and 0.15 dex respectively. The bottom panel shows the $\sim 2,000$
  stars for which the error on $T_{eff}$ is below 400K, the error on
  log~$g$ is below 0.5~dex and the signal-to-noise is at least 15. }
  \label{fig:GES}
\end{figure}

That there are indeed two distinct sequences in elemental abundances
for stars that probe the Galaxy several kiloparsecs from the Sun is
shown in Fig.~\ref{fig:GES}. These sequences are also distinct in
kinematics \citep{GES14,GK15}. This result will (presumably!) be
strengthened by the larger samples available from later data releases,
and by the addition of Gaia astrometric data. 

The near-IR capability of the APOGEE survey allows the study of stars
in the thin disk at lower latitudes than feasible in optical surveys,
and thus a larger range of radial coordinate for given range of
distances from the Sun. The high-resolution, high signal-to-noise
spectra ($R \sim 30,000$) provide precise estimates of the abundances
of individual elements, with internal errors in [$\alpha$/Fe] around
0.03~dex for red clump stars \citep[][see also Allende Prieto's
contribution to this conference]{Nid14}. Again, the APOGEE data reveal two
distinct sequences in elemental abundances, populated by stars with
distinct kinematics, as manifest in the locations of the stars with
respect to the Galactic mid-plane. The radial coverage of the APOGEE
survey further showed the thick disk (\lq high-alpha') sequence to be more centrally concentrated within the Galaxy than the thin disk (\lq low-alpha') sequence.  This inferred shorter scale-length for the thick disk would lead to a larger mass, for given local normalization. The spatial variations of the elemental abundance patterns in the Gaia-ESO survey are under investigation \citep{GK15}. 

The GALAH survey with the HERMES high-resolution spectrograph on the
AAT is obtaining elemental abundances and kinematics for a large
sample of stars in the thin disk, as described in this volume by Sarah
Martell, with a major goal to identify past star-formation events through chemical tagging. 

\subsubsection{Subaru Prime Focus Spectrograph Galactic Archaeology Survey}

The Subaru Prime Focus Spectrograph (PFS) project was presented in the
conference by Masahiro Takada, with an emphasis on the proposed extragalactic
surveys.  There will also be a Galactic Archaeology component, with science goals of constraining the nature of dark
matter through investigations of the density law in the Milky Way and
selected satellite galaxies, and of the evolutionary histories of each of the Milky Way and the Andromeda galaxy, including estimations of their merger histories.

 As described in \citet{Tak14}, the PFS project will take advantage of
the wide FoV ($\sim 1.4$~deg) of the Subaru 8\,m telescope, to feed a 3-arm
(blue, red, near IR) low-resolution ($R \sim 2000$) spectrograph with
over 2,000 fibers. These low-resolution spectra suffice for the
derivation of line-of-sight velocities and stellar chemical abundances (iron, carbon) with errors similar
to those of the SDSS SEGUE surveys. The Galactic Archaeology survey
will also utilize a medium resolution ($R \sim 5000$) mode for the red
arm. This medium resolution mode will enable the derivation of the
abundances of several of the alpha-elements, using the techniques
developed by Kirby and collaborators \citep{Kirby08,Kirby09}.

Given the location of the telescope in the Northern hemisphere,
targets in the Milky Way include the relatively understudied outer
disk, of particular interest in terms of constraining the importance
of internal secular effects such as stellar radial migration
\citep{SB02}, or \lq compression waves' and \lq breathing modes' of the
disk \citep{Wid14}. Standard cold dark matter dominated models of
galaxy formation predict that the outer regions form later and
accretion of high-angular momentum material - in the form of both
satellite systems and gas -- is expected at late times, such that the outer disk should be where significant substructure should be found.  The moderate
resolution mode will be used to complement the Gaia astrometric data
for stars that are brighter than $V \sim 20$, obtaining line-of-sight
velocities and alpha-abundances (and estimates of the stellar overall
metallicity, surface gravity and effective temperature).  Wider
volumes around known substructure in the Galaxy will also be targeted,
to provide better discrimination between external and internal
mechanisms for their creation. 

The PFS GA survey will also target fainter stars in the
low-resolution mode, including searches for substructure in metallicity-kinematic phase space. The distant giant stars will also be used in constraining the overall potential well of the Milky Way, and also allow us to determine
the edge of the Milky Way halo. 

The Andromeda galaxy (M31), the other large disk galaxy in the Local Group, will also be a major component of the (low-resolution) 
survey. We aim to determine the large-scale properties of M31, plus distinguish plausible tidal debris from substructure created from disturbances to M31 itself. This will facilitate analysis of the merger/accretion history. PFS will obtain 
spectroscopic metallicity estimates for
individual stars in M31, in contrast to photometric metallicities
\citep[e.g.][]{AF02,Gil14,Ibata14} or those obtained from stacking/co-adding
of spectra of many stars \citep[e.g.][]{raja06} that have previously
been used to determine metallicity distributions in M31.  PFS is
synergistic with HyperSuprimeCam and a dedicated narrow-band
(pre)imaging survey will be carried out, across the face of M31, to provide dwarf/giant separation (gravity sensitivity) so
that foreground (dwarf) stars in the Milky Way may be rejected from
the spectroscopic targets.  	\citet{Ibata14} discuss the complications and biases that can result when analyzing broad-band photometric data in isolation.

The wide field and high multiplexing capability of PFS allow study of
the nearest dwarf spheroidal satellite galaxies to out beyond their
nominal tidal radii, necessary for estimates of the total mass and to determine the amplitude of 
possible tidal effects. We will again carry out narrow-band
(pre)imaging with HyperSuprimeCam to identify likely foreground
contamination. The medium-resolution mode will be employed to obtain
alpha-abundances and line-of-sight velocities. The chemical abundances are critical to constrain baryonic feedback, especially the strong energy injection invoked in models that seek to modify the inner dark matter profile \citep[e.g.][]{Gov12}.

\section{Conclusions}

The several ongoing or imminent surveys of nearby stars  
(will) provide transformative datasets
to decipher how normal disk galaxies such as the Milky Way and M31
form and evolve. They will yield unique insight into the nature of
dark matter, by both derived density profiles and merger
histories. Contemporaneous high-redshift surveys are quantifying the
stellar populations and morphologies of galaxies at high lookback
times, and large, high-resolution simulations of structure formation
are allowing predictions of Galaxy formation in a cosmological
context, ready for testing. These three astrophysical approaches to
understanding our place in the Universe are complemented by the
(re-start of) the LHC, which should reveal \lq physics beyond the
Standard Model' that could produce a dark matter particle. Exciting
times indeed.

\acknowledgements I thank Ian Skillen and Scott Trager for
facilitating my participation and for their tireless energy making this
conference so enjoyable. I acknowledge NSF grant PHY-1066293 which
supports the Aspen Center for Physics and thank the Aspen Center for
Physics for providing a wonderful environment for physics while I
wrote this contribution. I acknowledge support from NSF grant
OIA-1124403. I thank Doroth\'ee Brauer for creating
Fig.~\ref{fig:RAVE_SEGUE}.

\bibliography{RWyse}

\begin{thebibliography}{}
\expandafter\ifx\csname natexlab\endcsname\relax\def\natexlab#1{#1}\fi
\expandafter\ifx\csname url\endcsname\relax
  \def\url#1{\texttt{#1}}\fi
\expandafter\ifx\csname urlprefix\endcsname\relax\def\urlprefix{URL }\fi
\providecommand{\eprint}[2][]{\url{#2}}

\bibitem[{{Adibekyan} et~al.(2013)}]{Adi13}
{Adibekyan}, V.~Z., et~al. 2013, \aap, 554, A44

\bibitem[{{Bahcall} \& {Soneira}(1980)}]{BS80}
{Bahcall}, J.~N., \& {Soneira}, R.~M. 1980, \apjs, 44, 73

\bibitem[{{Belokurov} et~al.(2006)}]{Bel06}
{Belokurov}, V., et~al. 2006, \apjl, 642, L137

\bibitem[{{Bensby} et~al.(2014){Bensby}, {Feltzing}, \& {Oey}}]{Bensby14}
{Bensby}, T., {Feltzing}, S., \& {Oey}, M.~S. 2014, \aap, 562, A71

\bibitem[{{Bond}(1970)}]{Bond70}
{Bond}, H.~E. 1970, \apjs, 22, 117

\bibitem[{{Bovy} et~al.(2012{\natexlab{a}}){Bovy}, {Rix}, \& {Hogg}}]{Bovy12b}
{Bovy}, J., {Rix}, H.-W., \& {Hogg}, D.~W. 2012{\natexlab{a}}, \apj, 751, 131

\bibitem[{{Bovy} et~al.(2012{\natexlab{b}}){Bovy}, {Rix}, {Liu}, {Hogg},
  {Beers}, \& {Lee}}]{Bovy12a}
{Bovy}, J., {Rix}, H.-W., {Liu}, C., {Hogg}, D.~W., {Beers}, T.~C., \& {Lee},
  Y.~S. 2012{\natexlab{b}}, \apj, 753, 148

\bibitem[{{Carney} et~al.(1989){Carney}, {Latham}, \& {Laird}}]{CLL89}
{Carney}, B.~W., {Latham}, D.~W., \& {Laird}, J.~B. 1989, \aj, 97, 423

\bibitem[{{Casagrande} et~al.(2011){Casagrande}, {Sch{\"o}nrich}, {Asplund},
  {Cassisi}, {Ram{\'{\i}}rez}, {Mel{\'e}ndez}, {Bensby}, \& {Feltzing}}]{Cas11}
{Casagrande}, L., {Sch{\"o}nrich}, R., {Asplund}, M., {Cassisi}, S.,
  {Ram{\'{\i}}rez}, I., {Mel{\'e}ndez}, J., {Bensby}, T., \& {Feltzing}, S.
  2011, \aap, 530, A138

\bibitem[{{Cheng} et~al.(2012)}]{cheng12a}
{Cheng}, J.~Y., et~al. 2012, \apj, 746, 149

\bibitem[{{Dotter} et~al.(2011){Dotter}, {Sarajedini}, \&
  {Anderson}}]{Dotter11}
{Dotter}, A., {Sarajedini}, A., \& {Anderson}, J. 2011, \apj, 738, 74

\bibitem[{{Eggen} et~al.(1962){Eggen}, {Lynden-Bell}, \& {Sandage}}]{Eggen62}
{Eggen}, O.~J., {Lynden-Bell}, D., \& {Sandage}, A.~R. 1962, \apj, 136, 748

\bibitem[{{Eisenstein} et~al.(2011)}]{Eis11}
{Eisenstein}, D.~J., et~al. 2011, \aj, 142, 72

\bibitem[{{Ferguson} et~al.(2002){Ferguson}, {Irwin}, {Ibata}, {Lewis}, \&
  {Tanvir}}]{AF02}
{Ferguson}, A.~M.~N., {Irwin}, M.~J., {Ibata}, R.~A., {Lewis}, G.~F., \&
  {Tanvir}, N.~R. 2002, \aj, 124, 1452

\bibitem[{{Fuhrmann}(2011)}]{Fuhr11}
{Fuhrmann}, K. 2011, \mnras, 414, 2893

\bibitem[{{Gilbert} et~al.(2014)}]{Gil14}
{Gilbert}, K.~M., et~al. 2014, \apj, 796, 76

\bibitem[{{Gilmore} \& {Reid}(1983)}]{GR83}
{Gilmore}, G., \& {Reid}, N. 1983, \mnras, 202, 1025

\bibitem[{{Gilmore} \& {Wyse}(1986)}]{GW86}
{Gilmore}, G., \& {Wyse}, R.~F.~G. 1986, Nature, 322, 806

\bibitem[{{Gilmore} \& {Wyse}(1998)}]{GW98}
--- 1998, \aj, 116, 748

\bibitem[{{Gilmore} et~al.(2012)}]{GES12}
{Gilmore}, G., et~al. 2012, The Messenger, 147, 25

\bibitem[{{Governato} et~al.(2012)}]{Gov12}
{Governato}, F., et~al. 2012, \mnras, 422, 1231

\bibitem[{{Guhathakurta} et~al.(2006)}]{raja06}
{Guhathakurta}, P., et~al. 2006, \aj, 131, 2497

\bibitem[{{Ibata} et~al.(1995){Ibata}, {Gilmore}, \& {Irwin}}]{IGI95}
{Ibata}, R.~A., {Gilmore}, G., \& {Irwin}, M.~J. 1995, \mnras, 277, 781

\bibitem[{{Ibata} et~al.(1997){Ibata}, {Wyse}, {Gilmore}, {Irwin}, \&
  {Suntzeff}}]{Ibata97}
{Ibata}, R.~A., {Wyse}, R.~F.~G., {Gilmore}, G., {Irwin}, M.~J., \& {Suntzeff},
  N.~B. 1997, \aj, 113, 634

\bibitem[{{Ibata} et~al.(2014)}]{Ibata14}
{Ibata}, R.~A., et~al. 2014, \apj, 780, 128

\bibitem[{{Jayaraman} et~al.(2013){Jayaraman}, {Gilmore}, {Wyse}, {Norris}, \&
  {Belokurov}}]{apoorva13}
{Jayaraman}, A., {Gilmore}, G., {Wyse}, R.~F.~G., {Norris}, J.~E., \&
  {Belokurov}, V. 2013, \mnras, 431, 930

\bibitem[{{Jones} \& {Wyse}(1983)}]{JW83}
{Jones}, B.~J.~T., \& {Wyse}, R.~F.~G. 1983, \aap, 120, 165

\bibitem[{{Kirby} et~al.(2009){Kirby}, {Guhathakurta}, {Bolte}, {Sneden}, \&
  {Geha}}]{Kirby09}
{Kirby}, E.~N., {Guhathakurta}, P., {Bolte}, M., {Sneden}, C., \& {Geha}, M.~C.
  2009, \apj, 705, 328

\bibitem[{{Kirby} et~al.(2008){Kirby}, {Guhathakurta}, \& {Sneden}}]{Kirby08}
{Kirby}, E.~N., {Guhathakurta}, P., \& {Sneden}, C. 2008, \apj, 682, 1217

\bibitem[{{Kordopatis} et~al.(2013)}]{RAVE13}
{Kordopatis}, G., et~al. 2013, \aj, 146, 134

\bibitem[{{Kordopatis} et~al.(2015)}]{GK15}
--- 2015, \aap, 582, A122

\bibitem[{{Leaman} et~al.(2013){Leaman}, {VandenBerg}, \& {Mendel}}]{Leaman13}
{Leaman}, R., {VandenBerg}, D.~A., \& {Mendel}, J.~T. 2013, \mnras, 436, 122

\bibitem[{{Lee} et~al.(2008)}]{SSP08}
{Lee}, Y.~S., et~al. 2008, \aj, 136, 2022

\bibitem[{{Lee} et~al.(2011)}]{SSP11}
--- 2011, \aj, 141, 90

\bibitem[{{Matteucci} \& {Chiappini}(2001)}]{Matt01}
{Matteucci}, F., \& {Chiappini}, C. 2001, New Astron Rev, 45, 567

\bibitem[{{Newberg} et~al.(2002)}]{New02}
{Newberg}, H.~J., et~al. 2002, \apj, 569, 245

\bibitem[{{Nidever} et~al.(2014){Nidever}, {Bovy}, {Bird}, {Andrews}, {Hayden},
  {Holtzman}, {Majewski}, \& et~al.}]{Nid14}
{Nidever}, D.~L., {Bovy}, J., {Bird}, J.~C., {Andrews}, B.~H., {Hayden}, M.,
  {Holtzman}, J., {Majewski}, S.~R., \& et~al. 2014, \apj, 796, 38

\bibitem[{{Nordstr{\"o}m} et~al.(2004)}]{Nord04}
{Nordstr{\"o}m}, B., et~al. 2004, \aap, 418, 989

\bibitem[{{Norris}(1987)}]{JEN87}
{Norris}, J. 1987, \apjl, 314, L39

\bibitem[{{Ratnatunga} \& {Freeman}(1989)}]{RF89}
{Ratnatunga}, K.~U., \& {Freeman}, K.~C. 1989, \apj, 339, 126

\bibitem[{{Recio-Blanco} et~al.(2014)}]{GES14}
{Recio-Blanco}, A., et~al. 2014, \aap, 567, A5

\bibitem[{{Searle} \& {Zinn}(1978)}]{SZ78}
{Searle}, L., \& {Zinn}, R. 1978, \apj, 225, 357

\bibitem[{{Sellwood} \& {Binney}(2002)}]{SB02}
{Sellwood}, J.~A., \& {Binney}, J.~J. 2002, \mnras, 336, 785

\bibitem[{{Snaith} et~al.(2015){Snaith}, {Haywood}, {Di Matteo}, {Lehnert},
  {Combes}, {Katz}, \& {G{\'o}mez}}]{Snaith15}
{Snaith}, O., {Haywood}, M., {Di Matteo}, P., {Lehnert}, M.~D., {Combes}, F.,
  {Katz}, D., \& {G{\'o}mez}, A. 2015, \aap, 578, A87

\bibitem[{{Steinmetz} et~al.(2006)}]{RAVE06}
{Steinmetz}, M., et~al. 2006, \aj, 132, 1645

\bibitem[{{Str{\"o}mgren}(1987)}]{Strom87}
{Str{\"o}mgren}, B. 1987, in The Galaxy, edited by G.~{Gilmore}, \&
  B.~{Carswell}, 229

\bibitem[{{Takada} et~al.(2014)}]{Tak14}
{Takada}, M., et~al. 2014, \pasj, 66, R1

\bibitem[{{Tinsley}(1975)}]{Tinsley75}
{Tinsley}, B.~M. 1975, \apj, 197, 159

\bibitem[{{Tinsley}(1979)}]{Tin79}
--- 1979, \apj, 229, 1046

\bibitem[{{Wallerstein}(1962)}]{Wall62}
{Wallerstein}, G. 1962, \apjs, 6, 407

\bibitem[{{Widrow} et~al.(2014){Widrow}, {Barber}, {Chequers}, \&
  {Cheng}}]{Wid14}
{Widrow}, L.~M., {Barber}, J., {Chequers}, M.~H., \& {Cheng}, E. 2014, \mnras,
  440, 1971

\bibitem[{{Wyse} \& {Gilmore}(1986)}]{WG86}
{Wyse}, R.~F.~G., \& {Gilmore}, G. 1986, \aj, 91, 855

\bibitem[{{Wyse} \& {Gilmore}(1992)}]{WG92}
--- 1992, \aj, 104, 144

\bibitem[{{Wyse} \& {Gilmore}(1993)}]{WG93}
--- 1993, in The Globular Cluster-Galaxy Connection, edited by G.~H. {Smith},
  \& J.~P. {Brodie}, vol.~48 of Astronomical Society of the Pacific Conference
  Series, 727

\bibitem[{{Wyse} \& {Gilmore}(1995)}]{WG95}
--- 1995, \aj, 110, 2771

\bibitem[{{Xu} et~al.(2015){Xu}, {Newberg}, {Carlin}, {Liu}, {Deng}, {Li},
  {Sch{\"o}nrich}, \& {Yanny}}]{Xu15}
{Xu}, Y., {Newberg}, H.~J., {Carlin}, J.~L., {Liu}, C., {Deng}, L., {Li}, J.,
  {Sch{\"o}nrich}, R., \& {Yanny}, B. 2015, \apj, 801, 105

\bibitem[{{Yanny} et~al.(2009)}]{Yan09}
{Yanny}, B., et~al. 2009, \aj, 137, 4377

\bibitem[{{Zinn}(1985)}]{Zinn85}
{Zinn}, R. 1985, \apj, 293, 424

\end{thebibliography}

 \end{document}